\documentclass[preprint,showpacs,preprintnumbers,amsmath,amssymb]{revtex4}

\usepackage{graphicx}
\usepackage{dcolumn}
\usepackage{bm}
\usepackage{textcomp} 

\begin{document}

\preprint{APS/123-QED}

\title{Quantitative Measurement of the Surface Charge Density}

\author{Florian Johann} \author{Elisabeth Soergel}
\email{soergel@uni-bonn.de} \affiliation{Institute of Physics,
  University of Bonn, Wegelerstra\ss e 8, 53115 Bonn, Germany}


\begin{abstract}
  We present a method of measuring the charge density on
  dielectric surfaces. Similar to electrostatic force microscopy we
  record the electrostatic interaction between the probe and the
  sample surface, but at large tip-sample distances. For
  calibration we use a pyroelectric sample which allows us to alter
  the surface charge density by a known amount via a controlled
  temperature change. For proof of principle we determined the surface charge density under ambient conditions of ferroelectric lithium niobate.
\end{abstract}


\maketitle

The measurement of charge densities on dielectric surfaces have been
investigated for more then fifty years~\cite{Gro45}. Initially
driven by the investigation of electrets~\cite{Gut48} the
determination of surface charge densities (SCD) has become a vivid
area of research, e.\,g.,~in semiconductors~\cite{Gar55}, liquid
crystals~\cite{Pag02} or hybrid materials~\cite{Lim03}. Quantitative
data of the SCD can provide material specific properties such as
information on lattice disruptions, an estimate on the degree
of alignment or a further insight into the interaction on the
micro-scale in compositions. In addition, SCD measurements allow
for the investigation of screening mechanisms on a quantitative level. Measuring the SCD is generally performed by means of non-contact electrical characterization techniques. In case of semiconductors on can take advantage of the surface photovoltage effect to determine the SCD~\cite{Sch01,Kne76,Tak97}. Because this technique relies on the generation of electron-hole pairs by illumination, it can not be applied to every material. A more general method to measure SCD is the vibrating probe technique which was introduced in the 1960s~\cite{Ree68} and later developed towards a scanning technique~\cite{Yut00}.  Here an electrode of known area is vibrated in the electric field generated by the charged surface to produce a signal which is proportional to the SCD.  Several methods based on capacitive probe for surface charge measurements have been
discussed since. They all suffer from the mandatory knowledge of the
capacity of the system which in general requires assumptions on the
geometrical shape of all participating components.

Electrostatic force microscopy (EFM) is a very powerful tool for
imaging charge distributions on dielectric surfaces~\cite{Ste88}. Because it is very sensitive, this technique allows the detection of single charges~\cite{Sch90} with a high lateral resolution in the nm-regime~\cite{Som99}. Although appropriate when used for mapping surface charge distributions, EFM, however, is a priori not applicable for quantitatively determining surface charge densities, again because it would require the exact knowledge of the shape of the probe.  When investigating charge distributions extending only some ten nanometers, the probe can be approximated by a sphere of radius~$r$ and an estimate on the magnitude of the surface charge can be obtained~\cite{Bri99}. This model, however, applies if and only if the dimensions of the charge distributions to be measured are of the order of the tip radius.

When investigating SCD the simple model approximating the probe by a
sphere does not apply any more since in general the spatial extend of the charging is large when compared to the size of the
probe. Explicitly the interaction between the cantilever and the
sample surface is no longer negligible. Thus for the quantitative
analysis of EFM signals obtained with homogeneously charged surfaces
one has to take into account the whole probe. This is insofar
difficult as the probe is geometrically complex. The cantilever for
instance can not be approximated by a beam of a certain length and
width because the edges, carrying a high charge density due to their small radius of curvature, must also be taken into account. To develop a model for the situation of a conducting probe above a charged surface that allows for quantitative conclusions is obviously very challenging.

In this contribution, we present a possibility to quantitatively
determine the surface charge density on dielectric surfaces by means
of a technique similar to EFM circumventing the above mentioned
difficulties. The key is to calibrate the entire system including the whole probe, and hence to avoid any assumptions on its exact
geometry. We were thus able to quantitatively determine the surface
charge density of ferroelectric lithium niobate under ambient
conditions.

The experiments were carried out with a commercial scanning force
microscope (SMENA from NT-MDT) with an alternating voltage of $V_{\rm ac} = 10$\,V$_{\rm pp}$ at a frequency of some 10\,kHz applied to
the conducting tip (DCP11 from NT-MDT). The electrostatic interaction between the periodically charged tip and the surface charges to be
detected leads to oscillations of the cantilever that are read out via a lock-in amplifier.  For a smooth tip-sample coarse approach we
upgraded the SFM with a piezoelectric transducer element that allows
for controlled distance changes over a range of 50\,\textmu m. Long
distance tip-sample approach curves were recorded with the help of
this additional piezo.

The sample was a pyroelectric crystal of large area ($15 \times
15$\,mm$^2$) and 500\,\textmu m thickness. Because of their high
values for the pyroelectric coefficients we chose lithium niobate
(${\rm d} P_s/{\rm d} T =-80 \times 10^{-6}\,$C/Km$^2$~\cite{Ges08})
and lithium tantalate (${\rm d} P_s/{\rm d} T =-190 \times
10^{-6}\,$C/Km$^2$~\cite{Gla68}). A further advantage of this choice
is that both materials are easily available with high, optically
polished surface quality.

\begin{figure}[p]
  \includegraphics{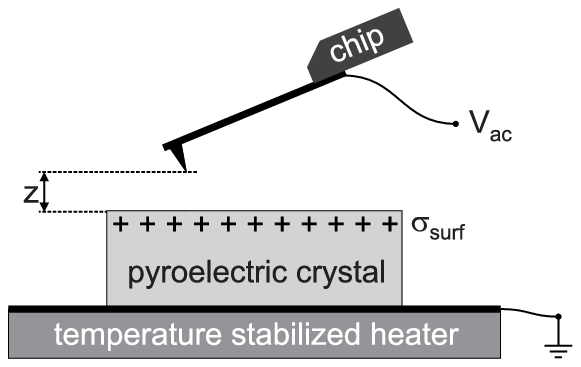}
\label{fig:joh01}
\caption{Schematics of the experimental setup. A scanning force microscope is operated as electrostatic force microscope with an alternating voltage $V_{\rm ac}$ applied to the tip. The tip-sample distance $z$ is several \textmu m. Calibration of the setup is performed with the help of a pyroelectric crystal which is  mounted on a stabilized oven. Upon a temperature change the surface charging $\sigma_{\rm surf}$ is altered in a controlled manner.
}
\end{figure}

The experimental setup is shown in Fig~\ref{fig:joh01}. The crystal is mounted with conductive thermopaste on top of a heater. The rear side of the crystal could therefore be grounded. According to need, we used either a temperature stabilized oven or a heating resistor for controlled temperature changes of the crystal. The temperature of the crystal was monitored with a PT100 resistance thermometer. The stabilized oven was slow, and temperature changes of 1\,K took several minutes, whereas with the heating resistor the crystals' temperature could be changed within few seconds.


Special care was taken to control the probe-sample
distance. Unfortunately the standard non-contact operation mode can in general not be utilized for surface charge density measurements. This is due to the fact that the electrostatic forces between tip and sample surface might by far exceed the van der Waals forces between
tip and sample used for non-contact operation. Measurements of the SCD must therefore be performed at large distance (several \textmu m) from the sample surface. For long-term measurements, e.\,g.~the temporal evolution of the compensation charging, it is  necessary to record data at the same probe-sample distance over several hours. As an additional challenge any contact between tip and sample surface must be avoided since this would inevitably influence any eventual compensation charges. The determination of the probe-sample distance can therefore not be based on the contact between tip and sample as a reference point. Obviously these requirements demand a very particular distance control which will be described below.

\begin{figure}[p]
  \includegraphics{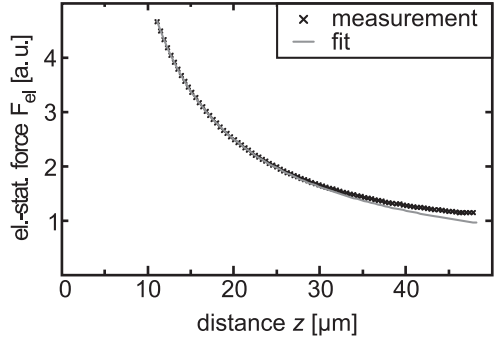}
  \label{fig:joh02}
  \caption{
Tip-sample approach curve above a $z$-cut LiNbO$_3$ crystal surface recording the electrostatic force $F_{\rm el}$ between probe and sample surface. Measurement and fit according to Eq.~\ref{Fit}.
}
\end{figure}

Let us review the experimental situation described above: a
periodically charged probe ($V\rm_{ac}=10\,V_{pp}$) at a distance $z$ of several microns of large-area sample ($> 100$\,mm$^2$) carrying a
homogeneous surface charge density $Q$. The electrostatic force
$F_{\rm el}$ acting on the probe in a homogeneous electric field is
proportional to the capacity $C$ of the system. Using the simple model of a conductive sphere opposed to a dielectric plane, $F_{\rm el}$ follows in the first approximation a $1/z$ dependence~\cite{Durand,Soergel}. Although $C$ is unknown, we can use this dependence to fit the data from the tip-sample approach curves according to the following empirical relation:
\begin{equation}\label{Fit}
  F_{\rm el}(z) \propto \frac{Q}{z} + B = \frac{Q}{z_0+\Delta z} + B
\end{equation}
where $Q$ is the SCD (to be calibrated in a later step)
and $B$ is a constant background. The distance $z$ can be separated into an unknown distance $z_0$ and a known variation $\Delta z$ performed by the piezoelectric transducer during the tip-sample approach curve. A typical curve on a $z$-cut LiNbO$_3$ crystal is shown in Fig.~\ref{fig:joh02}. From such a measurement, it is possible to determine all three parameters of Eq.~\ref{Fit} independently, in particular the tip-sample distance $z_0$ and the surface charge density $Q$. The deviations at longer distances are due to an increasing contribution of the cantilever to the signal with respect to that of the tip. Therefore the simple sphere model does not hold any more.  Since fitting is always performed with the steep part of the curve, the impact of this deviation is negligible.

\begin{figure}[p]
  \includegraphics{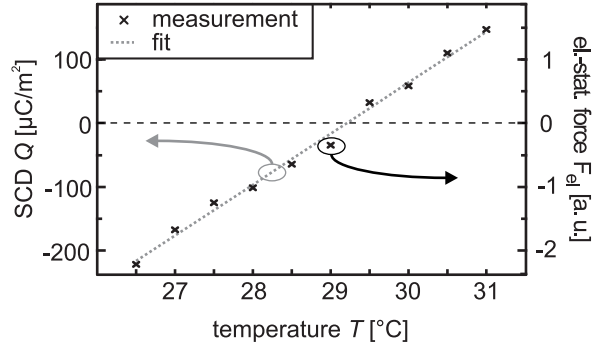}\label{fig:joh03}
  \caption{Calibration of the surface charge density (SCD) measurements with different temperature steps $\Delta T$ (starting each time from $T = 29^{\circ}$C) using a $z$-cut LiNbO$_3$ crystal.
}
\end{figure}

The last step for a quantitative determination of the SCD consists of the calibration of $F_{\rm el}$. For this purpose we took advantage of the pyroelectric effect, i.\,e.~the possibility to alter the SCD by a known amount $\Delta Q$ via a controlled change of the temperature $\Delta T$. The results of such measurements can be seen in Fig.~\ref{fig:joh03} where we changed the temperature by different steps $\Delta T <2.5^{\circ}$C starting with $T =29^{\circ}$C. Obviously this calibration procedure is very reliable within the chosen temperature range. Higher steps $\Delta T$ could not
be used because of the enormous SCD that built up lead to electrical
discharges via the tip. We furthermore changed the time $\Delta t$
over which $\Delta T$ happened from few seconds to several minutes,
however, no change in the calibration could be observed. From this we expect any external compensation or internal relaxation mechanisms to take place on a much longer time scale. To underline the reliability of our calibration procedure we also used a LiTaO$_3$ crystal which has a much larger pyroelectric coefficient. The results obtained were consistent within the precision to which the pyroelectric coefficients of the materials are known.

\begin{figure}[p]
  \includegraphics{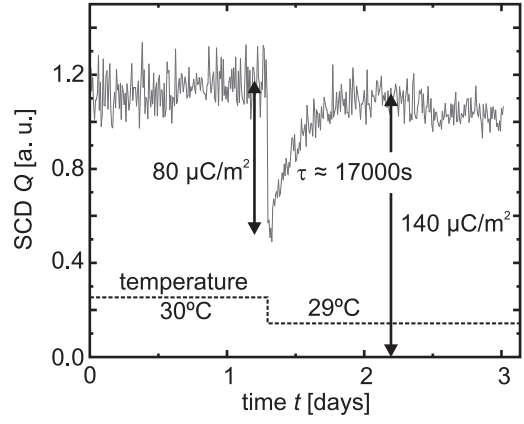}\label{fig:joh04}
  \caption{Time evolution of the real surface charge density (SCD) on a $z$-cut $\rm LiNbO_3$ sample over several days. The relaxation of the surface charging takes place with a time constant of $\tau = 17.000\,\rm s$.}
\end{figure}

Finally we investigated the long-term evolution of the SCD using a
single domain $z$-cut LiNbO$_3$ crystal. We therefore mounted the
crystal on the temperature stabilized oven, which was made very steady by a copper block. We recorded $Q$ over a timescale of three days, where we changed the temperature by 1$^{\circ}$C after half the time -- from $T_0 = 30^{\circ}$C to $T_1 = 29^{\circ}$C. Figure~\ref{fig:joh04} shows the result of the measurement. For every data point we took a tip-sample approach curve as the one shown in Fig.~\ref{fig:joh02} to correct for any drift of the distance during the measurement. From the measurement we could determine the relaxation time constant to be $\tau \approx 2 \times 10^4\,$s  and
$Q = 140$\,\textmu \rm C/m$^2$ for the reached long-term surface charge density.

The relaxation time constant $\tau$ for compensation of the
pyroelectric surface charging is given by~\cite{Ros00}
\begin{equation}\label{Relax}
  \tau = \frac{\varepsilon \cdot \varepsilon_0}{\sigma}
\end{equation}
where is $\varepsilon$ the relative permittivity of the crystal,
$\varepsilon_0$ is the permittivity of free space, and $\sigma$ is the bulk conductivity of the crystal. For congruent LiNbO$_3$ $\varepsilon = 28$ and $\sigma = 10^{-16}$ -- $10^{-18}\,\rm (\Omega cm)^{-1}$ therefore $\tau =10^3$ -- $10^5$\,s. Indeed our experimental result (Fig.~\ref{fig:joh04}) is in the range of this theoretical value. This is why we suppose that a change of the surface charge due to the pyroelectric effect is compensated by internal relaxation via bulk conductivity.


In order to verify our assumption we investigated iron-doped
LiNbO$_3$ crystals which exhibit a much higher dark conductivity~\cite{Bra08}. As expected, the relaxation time was
drastically shortened for higher doping levels. For  very low
doping of only 0.05\% the relaxation time $\tau$ could not be
determined accurately. We explain this by the fact that the controlled temperature change of the crystal is slow, of the order of minutes, therefore substantial compensation occurred already during heating or cooling.

Finally we want to comment on the measured value of $ Q =
140$\,\textmu \rm C/m$^2$ for the surface charge density on
congruently melt, undoped $z$-faced LiNbO$_3$ crystals. This value
signifies an attenuation of the spontaneous polarization of
0.7\,C/m$^2$ by a factor of $2\times 10^{-4}$~\cite{Wem68}. Different scenarios regarding the state of the surface charge of ferroelectric crystals have been considered and range from unscreened, partly or completely screened up to an over screened scenario~\cite{Alexe}. For the determination of the potential difference of oppositely oriented domains, different techniques that probe the electrostatic state of the sample surface such as electrostatic force microscopy (EFM) and scanning surface potential microscopy  have been employed (SSPM)~\cite{Bonnell}. Measurements, however, did not show any significant results~\cite{Scr05}. Although the value we measured might be influenced by the ambient conditions, it however gives the right order of magnitude for the surface charge density, as was shown by repetitive measurements. We also investigated etched surfaces, thereby excluding any drastic effect of the polishing procedure on the SCD.  Note that for an even more reliable value, controlled conditions (i.e. mainly humidity and more stable temperatures) are mandatory. It would be very instructive to repeat such measurements under ultra high vacuum conditions. This, however, is beyond our possibilities.

In conclusion we have demonstrated a straight forward measurement
scheme for quantitatively determining real surface charge densities on ferroelectric crystals. We exemplified our technique using a lithium niobate crystal, thereby revealing a compensation of the spontaneous polarization for the surface charge density by four orders of magnitude.

{\footnotesize {\bf Acknowledgements} Many thanks to Akos Hoffmann and
  Tobias Jungk for fruitful discussions.
  Financial support from the Deutsche Telekom AG is gratefully
  acknowledged.}


\newpage

\eject

\eject

\eject

\end{document}